# Adaptive Modulation and Coding and Cooperative ARQ in a Cognitive Radio System

J. Hwang, H. Saki*, and M. Shikh-Bahaei*
Centre for Telecommunications Research (CTR), King's College London, UK

*Abstract*—In this paper, a joint cross-layer design of adaptive modulation and coding (AMC) and cooperative automatic repeat request (C-ARQ) scheme is proposed for a secondary user in a shared-spectrum environment.

First, based on the statistical descriptions of the channel, closed-form expressions of the average spectral efficiency (SE) and the average packet loss rate (PLR) are presented. Then, the cross-layer scheme is designed, with the aim of maximizing the average SE while maintaining the average PLR under a prescribed level. An optimization problem is formed, and a sub-optimal solution is found: the target packet error rates (PER) for the secondary system channels are obtained and the corresponding sub-optimal AMC rate adaptation policy is derived based on the target PERs. Finally, the average SE and the average PLR performance of the proposed scheme are presented.

## I. INTRODUCTION

The fifth generation wireless network is expected to fulfill the objectives that have been set by regulatory bodies regarding the connection of all types of devices. Next generation networks should enable smart devices ranging from user equipments (UEs) and vehicles to sensors and actuators, to connect and communicate in the internet of things (IoT) [1]. Relay communications and cognitive radio (CR), or dynamic spectrum access (DSA), have received much attention as ways of achieving efficient utilization of radio spectrum [2, 3, 4]. In a CR network, unlicensed users (or secondary users (SU)) of a spectrum band are allowed to share the spectrum with the licensed users (primary users (PU)), as long as the quality of the PU transmissions is not impaired [5]. Among several models of spectrum sharing, the hierarchical access model is one of the most compatible models with the existing spectrum management methods [6, 7, 8]. In particular, in an underlay CR model, the SU is allowed to transmit data at the same time as the PU, with a constraint on the amount of transmission power of the SU so that interference on the PU is limited. Joint cross-layer design of AMC at the physical layer and ARQ at the data link layer (DLL) was considered [9, 10] as a way of improving spectrum efficiency (SE) while maintaining delay and error requirements [11]. In particular, [9] considered the use of the joint design for a SU system in a shared-spectrum environment; there, the statistics of the SU transmit power and of received channel quality were also found, in relation to the limits on the average and/or instantaneous interference on PU. The performance of the cross-layer design in [10] was investigated for a single point-to-point link between a secondary source and its destination. Closed-form expressions for the optimum retry-limit (RTL), overall loss rate and packet overflow drop rate are derived using queuing model in [12]. Cooperative communication (or relay-assisted communication) [13] exploits the broadcasting nature of wireless transmission to achieve diversity gain. A cross-layer approach which combines real-time RTL adaptation at medium access control (MAC) layer was developed in [14]. The ergodic capacity and outage probability performances of relay-assisted systems in spectrum sharing environments have been investigated in [15, 16]. In [17], a relay node was utilized for the ARQ operation at DLL, for a scheme known as cooperative ARQ (C-ARQ): when the channel quality between source and destination node is poor, the relay can participate in the transmission of packets provided that it has correctly received the packets from the source.

In this paper, we investigate the performance of the cross-layer combination of AMC and C-ARQ for secondary users under spectrum-sharing constraints. Namely, the overall average SE and average packet loss rate (PLR) are evaluated for a secondary system which employs a cooperative node for its ARQ re-transmissions. The two performance metrics are first presented as closed-form mathematical expressions, which are statistical averages obtained using the statistics of the channels' states. Then, the performance expressions are used in designing a rate adaptation policy for the physical layer; the aim of the adaptation policy is to achieve the maximum possible average SE while satisfying a constraint on the QoS in terms of the average PLR. It is found that designing a rate adaptation scheme translates into solving a non-linear optimization problem.

The performance of the secondary system employing the optimized rate adaptation and C-ARQ [18] is quantified for different levels of average and/or instantaneous interference on the primary receiver, and the constraints on the interference level are used in transmit power adaptation at the secondary nodes.

*This work was partially supported by the UK Engineering and Physical Sciences Research Council (EPSRC) grant numbers EP/P022723/1 and EP/P003486/1.

## II. SYSTEM MODEL

Consider a system where a primary transmitter (PT) and a primary destination (PD) node share a frequency band of bandwidth $B$ Hz with SUs, as in Figure 1.

The links ST-SD, SR-SD, and ST-SR are numbered as links 1, 2, and 3 with channel power gains $h_{s1}$, $h_{s2}$ and $h_{s3}$ respectively, and the interference channels have powers $h_{p1}$ (from ST to PD) and $h_{p2}$ (from SR to PD). Slow block fading scenario is considered, where the channel gains are assumed to be independent and identically distributed (i.i.d) stationary random variables which vary on a frame-by-frame basis with an associated probability distribution function (pdf).

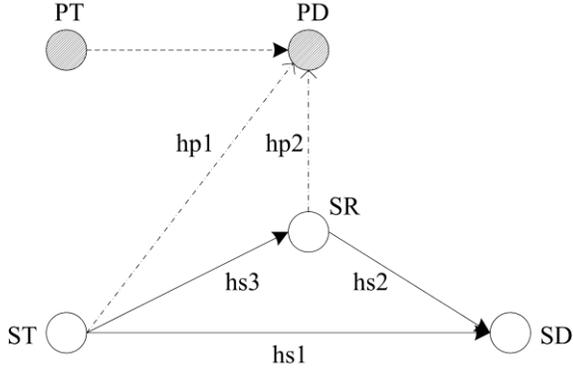

Figure 1. System Model of a secondary C-ARQ System

Here, Rayleigh fading model is used, hence pdf $f_{H_s}(h_{sv}) = \exp(-h_{sv})$ (for link $v = 1, 2, 3$) with unit mean (i.e. $E[h_{sv}] = E[h_{p1}] = E[h_{p2}] = 1$).

### A. Cooperative ARQ Scheme

At the DLL of ST, the stream of data is processed in the unit of a packet, which consists of $N_p$ bits. Using an ARQ scheme, when a negative acknowledgment (NACK) is received indicating that a packet was in error at destination node(s), the packet is allowed to be re-transmitted for the maximum of $N_r^{max}$ times.

In this paper, the cooperative ARQ scheme in [17] is considered, where a relay node (SR) (chosen through a given selection mechanism) assists transmission of data from ST to SD using decode-and-forward (DF) protocol. More specifically, when data is transmitted from ST to SD, SR also listens and tries to decode the packets. If the received packet in SD is in error and SR has successfully decoded the packet, it forwards the packet to the SD, and sends an ACK message to ST to prevent it from re-transmitting the packet. In other words, ST would repeat transmission until a packet is received correctly either at SD or SR, or the total number of re-transmission reaches $N_r^{max}$. Let $l$ and $k$ be the numbers of re-transmissions from ST and SR respectively, where $0 \leq l + k \leq N_r^{max}$. The case where $l = 0$ and $k = 0$ means that ST to SD transmission succeeded at the first attempt and no ARQ re-transmission would be employed. A packet would be declared lost if it is received in error at SD after $N_r^{max} + 1$ transmissions, and a given limit $P_{loss}$ is set on the probability of packet loss for the overall cooperative system.

### B. Power Allocation and AMC at the Physical Layer

At the physical layer, data bits in each packet structure would be mapped to symbols with modulation, and the resulting symbols are grouped into a frame structure with $N_s$ symbols in total (this includes pilot symbols for channel estimation purpose). Assuming ideal Nyquist pulses, the symbol duration $T_s$ is equal to $1/B$. At the start of each frame, transmit powers $P_1$ (at the ST) and $P_2$ (at the SR) are determined subject to the constraints placed on the levels of average and instantaneous interference at PD. That is, the interference channels ST-PD, and SR-PD should have 1) the average received SNR below a given limit $Q_a$ and/or 2) instantaneous SNR below a given limit $Q_{ins}$. In this paper, it is assumed that $Q_{ins} = \eta \cdot Q_a$, where $\eta$ may take different values such as 1, 1.5 or infinity (i.e. $\eta = \infty$). Setting $\eta = \infty$ (hence $Q_{ins} = \infty$) corresponds to the case of having no limit on the amount of instantaneous interference the SD or the SR inflict on the PR.

Representing this mathematically, the optimal transmit power $P_v$ is determined such that:

$$E[P_v h_{pv}] \leq Q_a \text{ and } P_v \cdot h_{pv} \leq Q_{ins}. \quad (1)$$

Then, the transmit power of link $v$ is:

$$P_v = \min\left\{Q_a, \frac{\eta Q_a}{h_{pv}}\right\} = \begin{cases} Q_a, & \text{if } \eta = \infty \\ \min\left\{Q_a, \frac{\eta Q_a}{h_{pv}}\right\}, & \text{if } \eta \neq \infty \end{cases} \quad (2)$$

When $v = 3$, $P_3$ is equal to $P_1$ since the transmitter of link 3 is ST. With transmit power $P_v$, the corresponding received SNR over link $v$ is $\gamma_v = P_v h_{sv}/I$, where $I$ is the sum of the power spectral densities of additive white Gaussian noise (AWGN) and other interferences at the receiving node.

For AMC, M-QAM modulation with order $M$ and convolutional code rate $R_c$ are used. The combined pair ($M_n$ and $R_{c,n}$) corresponds to *transmission mode* $n$, which is chosen from a given set of $N$ pairs (i.e. $n = 1, 2, ... N$). In the discrete-rate AMC operation, the range of all possible values of received SNR is divided into $N + 1$ regions by a set of SNR threshold values $\{\rho_{t1,n}, n = 0, 1, ..., N\}$, where $\rho_{t1,0} = 0$ and $\rho_{t1,N+1}$ is considered to be infinity [19]. Assuming perfect channel estimation at the receiver, when the received SNR $\gamma_1$ at the SD falls within the region $\rho_{t1,n} \leq \gamma_1 < \rho_{t1,n+1}$, transmission mode $n$ is chosen. This choice is fed back to the ST as the channel state information (CSI), via a fast and error-free feedback channel. If $\rho_{t1,0} \leq \gamma_1 < \rho_{t1,1}$, outage is declared and no transmission occurs. Transmission mode $m$ for the SR is chosen by a similar operation with the set of thresholds $\{\rho_{t2,n}\}$. For the sake of convenience, it is assumed that the ST and the SR use the same set of AMC modes; i.e. $\{n\} = \{m\}$ and $\rho_{t1,n} = \rho_{t2,n}$, $n = 1, 2, \cdots, N$.

At transmission mode $n$, transmission rate $r_n$ is defined as the amount of the transmitted information bits per symbol, and can be obtained as $R_{c,n} \cdot \log_2 M_n$ (bits/symbol).

$$\eta_s = \sum_{l=0}^{N_r^{max}} \sum_{k=0}^{N_r^{max}-l} \sum_{n_0=0}^{N} \cdots \sum_{n_l=0}^{N} \sum_{m_1=0}^{N} \cdots \sum_{m_k=0}^{N} \frac{\prod_{i=0}^{l-1} \overline{PER}_{1,n_i} \overline{PER}_{3,n_i}}{\Omega(\boldsymbol{n}_l, \boldsymbol{m}_k)}$$

$$\times \left\{ (1 - \overline{PER}_{1,n_l}) I_{\{k=0\}} + \overline{PER}_{1,n_l}(1 - \overline{PER}_{3,n_l})(1 - \overline{PER}_{2,m_k}) \left( \prod_{j=1}^{k-1} \overline{PER}_{2,m_j} \right) I_{\{k \geq 1\}} \right\} \prod_{i=0}^{l} \Pr(n_i) \prod_{j=1}^{k} \Pr(m_j) \quad (3)$$

$$\overline{PLR}_s = \sum_{l=0}^{N_r^{max}} \left( \overline{PER}_{(1,3)} \right)^l \times \left\{ \overline{PER}_1 I_{\{l=N_r^{max}\}} + \left( \overline{PER}_1 - \overline{PER}_{(1,3)} \right) \left( \overline{PER}_2 \right)^{N_r^{max}-l} I_{\{l<N_r^{max}\}} \right\} \quad (4)$$

### III. PERFORMANCE METRICS

In this section, the closed-form expressions of the overall average SE and the average PLR for the secondary system are derived statistically.

The pdf of the received SNR over a single secondary link $v$ was found in [9]:

$$f_{\Gamma_{sv}}(\gamma_{sv}) = \int_0^{Q_a} f_X(P_v) \cdot \frac{I}{P_v} f_{H_s}\left(\frac{\gamma_{sv} I}{P_v}\right) dP_v$$

$$+ \frac{I}{Q_a} f_{H_s}\left(\frac{\gamma_{sv} I}{Q_a}\right) \int_{Q_a}^{\infty} f_X(x) \, dx, \text{ where } X = \frac{Q_{ins}}{h_{pv}}. \quad (5)$$

Then, the probability of transmission mode $n_i$ being chosen by the ST at the $i^{th}$ transmission of a packet and the probability of transmission mode $m_j$ being chosen by the SR at the $j^{th}$ transmission of a packet, can be obtained using (5):

$$\Pr(n_i) = \int_{\rho_{t1,n_i}}^{\rho_{t1,n_i+1}} f_{\Gamma_{s1}}(\gamma_{s1}) \, d\gamma_{s1},$$

$$\Pr(m_j) = \int_{\rho_{t2,m_j}}^{\rho_{t2,m_j+1}} f_{\Gamma_{s2}}(\gamma_{s2}) \, d\gamma_{s2}. \quad (6)$$

The instantaneous packet error rate (PER) was closely approximated by a simple exponential equation in [20]; it relates the PER to the instantaneous received SNR $\gamma$ with constants $a_n$ and $g_n$:

$$PER_n(\gamma) = \begin{cases} 1, & 0 < \gamma < \rho_{t,n} \\ a_n \exp(-g_n \gamma), & \rho_{t,n} \leq \gamma < \rho_{t,n+1} \end{cases} \quad (7)$$

Using this information, the average PER links 1, 2, and 3 at a given transmission are:

$$\overline{PER}_{1,n_i} = \frac{1}{\Pr(n_i)} \int_{\rho_{t1,n_i}}^{\rho_{t1,n_i+1}} PER_{n_i}(\gamma_1) f_{\Gamma_1}(\gamma_1) \, d\gamma_1 \quad (8)$$

$$\overline{PER}_{2,m_j} = \frac{1}{\Pr(m_j)} \int_{\rho_{t2,m_j}}^{\rho_{t2,m_j+1}} PER_{m_j}(\gamma_2) f_{\Gamma_2}(\gamma_2) \, d\gamma_2 \quad (9)$$

$$\overline{PER}_{3,n_i} = \int_{\rho_{t1,n_i}}^{\rho_{t1,n_i+1}} PER_{n_i}(\gamma_3) f_{\Gamma_3}(\gamma_3) \, d\gamma_3. \quad (10)$$

These expressions were used in [17] to derive the average SE $\eta_s$ (in bits/sec/Hz) and the average PLR, presented in (3) and (4), for the overall secondary system.

The average SE was defined as the average number of accepted bits per symbol. $\boldsymbol{n}_l = \{n_i\}, i = 0,1, \dots, l$, $\boldsymbol{m}_k = \{m_j\}, j = 1,2, \dots, k$, and $I_{\{u\}} = 1$ if $u$ is true and 0 otherwise. $\Omega(\boldsymbol{n}_l, \boldsymbol{m}_k)$ is defined as $\sum_{i=0}^{l} 1/r_{n_i} + \sum_{j=0}^{k} 1/r_{m_j}$, and the product operator terms ($\prod \cdots$) take the value of 1 when the upper index is smaller than the lower index.

In (4), $\overline{PER}_{(1,3)}$ is the average probability of a packet being received in error at both SR and SD, and is defined as $\sum_{n=1}^{N} \overline{PER}_{1,n} \overline{PER}_{3,n} \Pr(n)$, while $\overline{PER}_1 = \sum_{n=1}^{N} \overline{PER}_{1,n} \Pr(n)$ is the average PER over link 1.

The case of $N_r^{max} = 0$ is equivalent to using no ARQ and no relay nodes, in which case the spectral efficiency expression (3) reduces to:

$$\eta_s^{AMC-ONLY} = r_n \Pr(n) \times (1 - \overline{PER}_{1,n}), \quad (11)$$

and the average PLR from (4) is reduced to:

$$\overline{PLR}_s^{AMC-ONLY} = \overline{PER}_1 = \sum_{n=1}^{N} \overline{PER}_{1,n} \Pr(n) \quad (12)$$

### IV. RATE ADAPTATION POLICY

An optimal rate adaptation policy for the secondary C-ARQ system can be designed such that the average SE is maximized subject to an average PLR constraint. For this purpose, (3) and (4) are used to form a constrained nonlinear optimization problem; the solution of the problem, $(\rho_{t_1}^*, \rho_{t_2}^*)$, is the set of SNR threshold values to be used for AMC (rate adaptation) operation:

$$\max_{\rho_{t_v,N} \geq \dots \geq \rho_{t_v} \geq \rho_{t_v,0}=0, \; v=1,2} \eta_s(\rho_{t_1}, \rho_{t_2}),$$

$$\text{subject to } \overline{PLR}(\rho_{t_1}, \rho_{t_2}) \leq P_{loss} \quad (13)$$

Considering the formulation of $\eta_s$ and $\overline{PLR}_s$ in (3) and (4), finding the optimum solution to this problem would be complex. Therefore, a sub-optimal solution is sought, by setting $\overline{PER}_{1,n}$ and $\overline{PER}_{2,m}$ to fixed values of target PERs, $P_{t1}$ and $P_{t2}$ respectively, and solving a new optimization problem;

$$\max_{0 < Pt_1, Pt_2 \leq 1} \eta_s(Pt_1, Pt_2)$$

$$\text{subject to } \overline{PLR}(Pt_1, Pt_2) \leq P_{loss}. \quad (14)$$

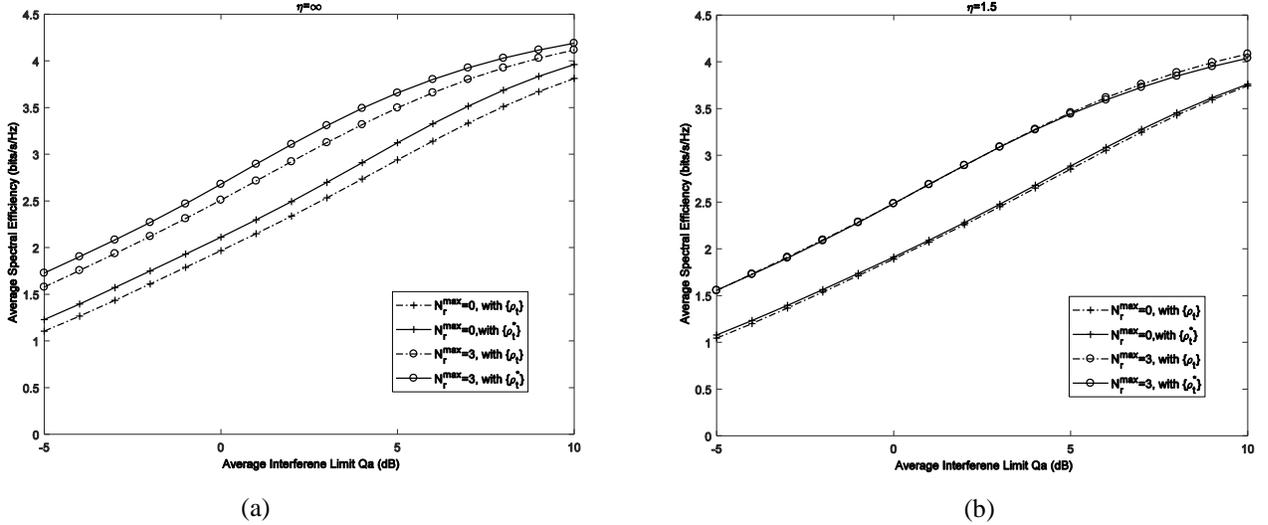

Figure 2. average SE (a) when $\eta = \infty$ and (b) when $\eta = 1.5$, ($I = -15$ dB, $P_{loss} = 10^{-4}$)

For solving the above problem, techniques such as the barrier function method (or the interior penalty function method) [21, 22] can be used, where the constrained problem such as (14) is transformed to an unconstrained optimization problem:

$$\max_{0<Pt_1,Pt_2\leq 1} T(\mu, Pt_1, Pt_2)$$
$$= \eta_s(Pt_1, Pt_2) - \mu\alpha(Pt_1, Pt_2). \quad (15)$$

In (15), the penalty parameter $\mu \geq 0$, and $\alpha(Pt_1, Pt_2)$ is defined as $-\log(P_{loss} - \overline{PLR}(Pt_1, Pt_2))$, so that when the values of $(Pt_1, Pt_2)$ cause violation of the constraint, $\alpha(Pt_1, Pt_2)$ would converge toward infinity. In this way, the feasibility of the solution is always fulfilled. To obtain the optimal solution $(P_{t1}^*, P_{t2}^*)$, (15) is solved for $k (= 1,2,3,...)$ iterations, where $\mu$ is decremented at each iteration. The solution is found when there is no improvement on the value of $T(\mu, Pt_1, Pt_2)$ as $k$ increases. That is, the solution $(P_{t1}^*, P_{t2}^*)$ is considered to be found when $\mu\alpha(Pt_1, Pt_2)$ is 0. In practice, the optimization iteration may be stopped when $\mu\alpha(Pt_1, Pt_2)$ is less than or equal to a very small value, e.g. $10^{-6}$.

$Pt_1^*$ and $Pt_2^*$ are the optimal target values for the average PER of link 1 and link 2 respectively. The sets of optimal SNR thresholds for the AMC operation at ST and SR, $\{\rho_{t1}^*\}$ and $\{\rho_{t2}^*\}$, can then be found by finding the solution $x$ of the following problem, for each transmission mode $n = 1,2,...,N$:

$$\int_x^{\rho_{tv,n+1}^*} (PER_n(\gamma) - P_{tv}^*) \cdot f_{\Gamma_1}(\gamma)\, d\gamma = 0, (v = 1,2) \quad (16)$$

where $\rho_{tv,0}^* = 0$ and $\rho_{tv,N+1}^* = \infty$.

## V. RESULTS AND DISCUSSIONS

In this section, numerical results are presented to illustrate the performance of a secondary system employing C-ARQ and AMC in a CR environment. For the choice of $(M_n, R_{c,n})$, the AMC table II (TM2) in [20] is used, where the total number of available transmission modes, $N$, is six.

First, performance of the optimized rate adaptation policy from (16) is compared with the performance of the AMC and T-ARQ combination in [9], with and without a requirement on the instantaneous received interference at PR (i.e. for $\eta = \infty$ and $\eta = 1.5$ ). In [9], the SNR thresholds $\rho_{t1,n}$ ($n = 1,2,...,N) \in \{\rho_{t1}\}$ for rate adaptation (AMC operation) at ST was simply set to satisfy a requirement on the target PER, $PER_{tgt}$. $PER_{tgt}$ was determined in the following way: since a packet is declared as 'lost' when it is received in error after $N_r^{max} + 1$ transmissions, it can be deduced that the instantaneous PER at the physical layer $PER_{PHY}$ should satisfy $(PER_{PHY})^{N_r^{max}+1} \leq P_{loss}$, and hence $PER_{PHY} \leq P_{loss}^{1/(N_r^{max}+1)}$. Then, $P_{loss}^{1/(N_r^{max}+1)}$ was defined as $PER_{tgt}$, which was used in setting $\rho_{t1,n}$ such that $\rho_{t1,n}$ is equal to $-\frac{1}{g_n}\log\left(\frac{PER_{tgt}}{a_n}\right)$. In contrast, the rate adaptation policy proposed in this paper uses a set of optimized SNR thresholds $\rho_{tv,n}^*$, with which the average SE of the overall secondary system is at its maximum while the average PLR is below the given target $P_{loss}$.

From Fig.2 (a), improvement on the average SE from the optimized rate adaptation policy can be seen for the case where $\eta = \infty$ i.e. when no limit is placed on the instantaneous interference on the PR. However, from Fig. 2 (b), when there is an extra limit on the instantaneous interference, it is seen that using the optimized rate adaptation design does not warrant better average SE performance. This can be inferred from the fact that the allocation of transmit powers $P_v^*$ ($v = 1,2$) from (2) depend on the value of $\eta$: When the transmit power $P_v^*$ at ST or SR is constrained by only the average interference limit, $P_v^*$ is always equal to $Q_a$. On the other hand, with the additional constraint on the instantaneous interference at PR, there is more likelihood of $P_v^*$ being lower than $Q_a$ when $(Q_{ins}/h_{p1}) < Q_a$ (i.e. $\eta < h_{pv}$). In this case, the corresponding received SNR $\gamma_v = P_v^* h_{sv}/I$ of the secondary channels would suffer in turn, and lower transmission rate would be selected from AMC.

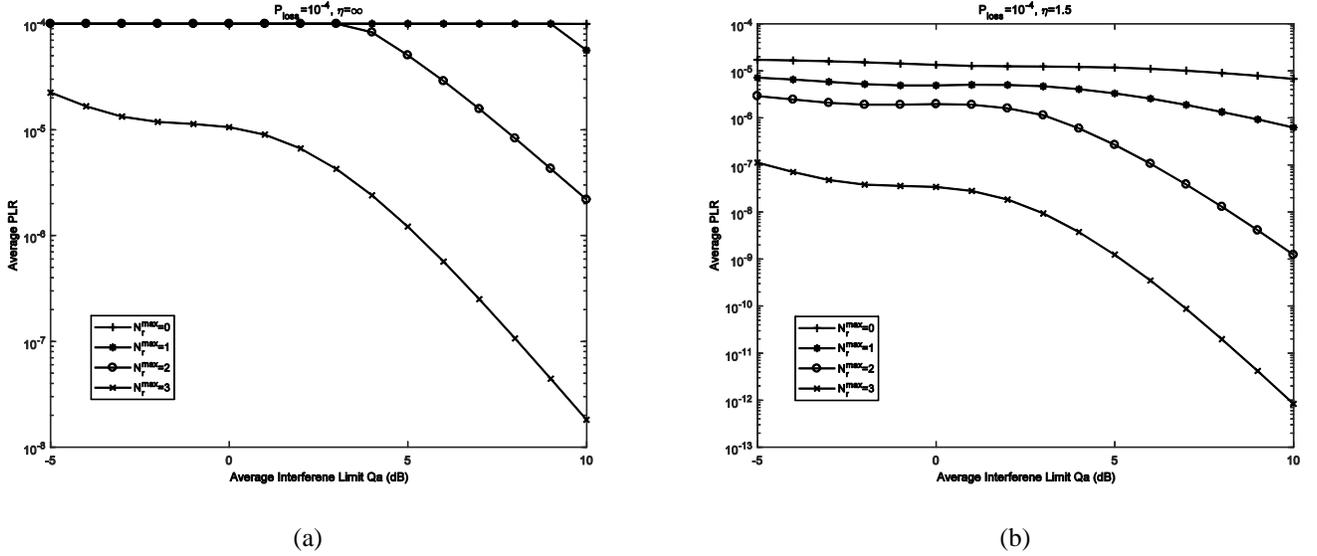

Figure 3. Average PLR against the average interference limit $Q_a$. ($I = -15$ dB)

Using the C-ARQ scheme, packets may be re-transmitted by either the ST or the SR, when the SR has correctly decoded the packet. The total number of transmissions for a packet is bounded by $N_r^{max} + 1$. Fig. 3 shows the variation of the average PLR as $N_r^{max}$ increases, for both $\eta = \infty$ and $\eta = 1.5$. It is observed that while the increase in $N_r^{max}$ does not improve the average SE, the average PLR performance is significantly lowered, as there would be more likelihood of a packet being correctly received with extra transmissions.

VI. CONCLUSION

The performance of relay-assisted secondary transmission in a shared spectrum environment is investigated, where a C-ARQ scheme at the DLL and AMC the physical layer are jointly considered. A rate adaptation policy is designed, such that the average SE is maximized while satisfying the QoS requirement at the DLL in terms of the average PLR. Improvements in the average SE and the average PLR performances are seen with the new rate adaptation policy.